\documentclass[aps,prl,twocolumn]{revtex4}

\usepackage{amsmath,amsthm,amsfonts,xspace,verbatim,graphicx}

\begin{document}
% The following line represents operators as roman text.  Another possibility is boldface.
%\newcommand{\op}[1]{{\bf #1 }}
\newcommand{\op}[1]{#1}
\newcommand{\ket}[1]{\left| #1 \right\rangle}
\newcommand{\bra}[1]{\left\langle #1 \right|}
\newcommand{\braket}[2]{\left\langle #1 | #2 \right\rangle}
\newcommand{\braopket}[3]{\bra{#1}#2\ket{#3}}
\newcommand{\proj}[1]{| #1\rangle\!\langle #1 |}
\newcommand{\expect}[1]{\left\langle#1\right\rangle}
\newcommand{\Entropy}{H}
\newcommand{\KL}[2]{S\left(#1\|#2\right)}
\newcommand{\Tr}{\mathrm{Tr}}
\newcommand{\Rho}{P}
\def\Id{1\!\mathrm{l}}
\newcommand{\cM}{\mathcal{M}}
\newcommand{\cR}{\mathcal{R}}
\newcommand{\cA}{\mathcal{A}}
\newcommand{\cL}{\mathcal{L}}
\newcommand{\reals}{\mathbb{R}}
\newcommand{\grad}{\nabla}
\newcommand{\rhohat}{\hat{\op\rho}}
\newcommand{\rhoMLE}{\rhohat_\mathrm{MLE}}
\newcommand{\rhoHMLE}{\rhohat_\mathrm{H}}
\newcommand{\rhotomo}{\rhohat_\mathrm{tomo}}
\newcommand{\rhotrue}{\rho}
% The following line represents vectors as boldface.  Another possibility is via \vec{}.
\newcommand{\pvec}[1]{{\bf #1}}
\newcommand{\pMLE}{\pvec{p}_{\mathrm{MLE}}}
\newcommand{\pHMLE}{\pvec{p}_{\mathrm{H}}}
\newcommand{\diff}{\mathrm{d}\!}
\newcommand{\pdiff}[2]{\frac{\partial #1}{\partial #2}}
\def\FCW{1.0\columnwidth}
\def\HCW{0.55\columnwidth}
\def\TPW{0.33\textwidth}

\graphicspath{{Figures/}}

\title{Hedged Maximum Likelihood Estimation}

\author{Robin Blume-Kohout}
\affiliation{Perimeter Institute for Theoretical Physics}
\email{robin@blumekohout.com}

\begin{abstract}
This paper proposes and analyzes a new method for quantum state estimation, called hedged maximum likelihood (HMLE).  HMLE is a quantum version of Lidstone's Law, also known as the ``add $\beta$'' rule.  A straightforward modification of maximum likelihood estimation (MLE), it can be used as a plugin replacement for MLE.  The HMLE estimate is a strictly positive density matrix, slightly less likely than the ML estimate, but with much better behavior for predictive tasks.  Single-qubit numerics indicate that HMLE beats MLE, according to several metrics, for nearly all ``true'' states.  For nearly-pure states, MLE does slightly better, but neither method is optimal.
\end{abstract}

\maketitle

Quantum state estimation is a basic task in quantum information science \cite{NielsenBook00}, simple to describe but hard to do right.  The estimator gets $N$ independently and identically prepared (i.i.d.) quantum systems, performs measurements on them, analyzes the data, and reports a single-system density matrix $\rhohat$.  The goal is to report the most ``accurate'' answer possible.  There is room for substantial debate over what this means, which the present paper will avoid by adopting three common assumptions: (i) we are concerned with $N$ copies of an unknown ``true'' state $\rhotrue$; (ii) the goal is to get $\rhohat$ as close as possible to $\rhotrue$, according to some metric $d(\rhotrue,\rhohat)$; and (iii) measurement outcomes are intrinsically random, and we are concerned with \emph{average} (over measurement outcomes, not over $\rhotrue$) performance.  I will not consider how to choose a measurement, seeking instead a protocol that works well for all measurements.

Maximum likelihood estimation (MLE) \cite{HradilPRA97,JamesPRA01,HradilLNP04, HaeffnerNature05}, the most common protocol, tends to report rank-deficient estimates with zero eigenvalues \cite{RBK06b}.  Those eigenvalues represent probabilities.  Assigning a zero probability indicates extraordinary confidence -- confidence that the data do not support.  For predictive purposes, this ``zero eigenvalue problem'' can be be disastrous in practice.

This paper suggests an alternative, \emph{hedged maximum likelihood}.  HMLE is a simple modification of MLE that can be used as a plug-in substitute for it.  The modification consists, in its entirety, of the following rule.  {\bf Replace the standard likelihood function $\cL(\rho) = \Pr(\mathrm{observed\ data}|\rho)$ with the product of $\cL(\rho)$ and a ``hedging function''
\begin{equation}\label{eq:HedgingFunction}
h(\rho) = \det(\rho)^\beta,
\end{equation}
where $\det(\cdot)$ is the determinant, and $\beta\approx\frac12$ is a positive constant chosen at the estimator's discretion.}  The rest of this Letter is devoted to explaining, deriving, and analyzing this procedure.

%\section{Classical hedging: the ``add-$\beta$'' rule}

\textbf{Background:}  HMLE is motivated by a rule for estimating classical probabilities called \emph{Lidstone's Law} \cite{LidstoneTFA20,Ristad95} -- or, more colloquially, ``add $\beta$''.  Suppose we have observed $N$ samples from an unknown i.i.d. distribution $\pvec{p} = \{p_1\ldots p_K\}$, and have seen $n_k$ ``$k$''s. What probabilities $\hat{\pvec{p}}$ should we assign for the next sample?  The likelihood, $\cL(\pvec{p}) = \prod_k{p_k^{n_k}}$, is maximized by the natural and obvious estimate
\begin{equation}\label{eq:NaiveRule}
\hat{p}_k = \frac{n_k}{N}.
\end{equation}
This can be disastrous in practice.  Suppose some letter $k$ has not yet been observed, so $n_k=0$, and MLE assigns $\hat{p}_k=0$.  This is fine if $p_k$ really is zero, but it's equally plausible that $p_k$ is positive but small.  If it is, the consequences of this error depend on what the estimate is used for.  They are catastrophic when the estimate is used for \emph{predictive} tasks, such as data compression or gambling \cite{CoverBook91,XieIEEE00}. 

Compression and gambling define operational interpretations of $\hat{\pvec{p}}$, and identify \emph{relative entropy} as a measure of error:
\begin{equation}
D(\pvec{p}||\pvec{\hat{p}}) = \sum_k{p_k\left(\log p_k - \log \hat{p}_k\right)}.
\end{equation}
A gambler maximizes his bankroll's expected growth rate by gambling a fraction $\hat{p}_k$ of it on outcome ``$k$'', and a compressor gets optimal performance by replacing ``$k$'' with a codeword of length $-\log\hat{p}_k$.  If the true probabilities are $\pvec{p}$ and the estimate is $\hat{\pvec{p}}$, then the gambler's wealth grows as $\$(n) = \$(0)e^{n(\mathrm{const}-H(\pvec{p})-D(\pvec{p}||\pvec{\hat{p}}))}$, where $H(\pvec{p}) \equiv -\sum_k{p_k\log p_k}$ is the entropy of $\pvec{p}$.  Similarly, the length of the compressor's compressed string grows as $L = n\left[H(\pvec{p}) + D(\pvec{p}||\pvec{\hat{p}})\right]$.  In both cases, $H(\pvec{p})$ is the unavoidable cost of $\pvec{p}$'s randomness, while $D(\pvec{p}||\pvec{\hat{p}})$ is the \emph{additional} cost of estimating it incorrectly.  Setting $\hat{p}_k=0$ thus implies extreme strategies for gambling (bet the entire bankroll against ``$k$'') and data compression (map ``$k$'' to an infinitely long codeword).  Either way, if the next letter is ``$k$'', the consequences are disastrous. 

``Add $\beta$'' avoids these catastrophes by hedging against as-yet-unseen possibilities.  It assigns probabilities
\begin{equation}\label{eq:AddBetaRule}
\hat{p}_k = \frac{n_k+\beta}{N+K\beta}.
\end{equation}
The lowest probability that can be assigned is $\frac{\beta}{N+K\beta} \approx \frac{\beta}{N}$.  Like Eq. \ref{eq:NaiveRule}, this rule has a statistical derivation.  It is the Bayes estimator (i.e., it minimizes expected cost) for a relative entropy cost function and a Dirichlet-$\beta$ prior
\begin{equation}\label{eq:DirichletPrior}
P_0(\pvec{p})\diff\pvec{p} \propto \prod_k{p_k^{\beta-1}\diff p_k}.
\end{equation}
Common examples of Dirichlet priors include the ``flat'' Lebesgue measure ($\beta=1$), and Jeffreys' prior ($\beta=\frac12$).  Given any prior, we can minimize expected relative entropy by: (1) updating the prior to a posterior via Bayes' Rule; and (2) reporting its mean value.  For the Dirichlet-$\beta$ prior, this gives the ``add $\beta$'' rule.

The ``add $\beta$'' rule is \emph{not} intrinsically Bayesian, however.  A n\"aive estimator following Eq. \ref{eq:NaiveRule} can simulate it by adding $\beta$ dummy observations of each letter $k$.  This yields new frequencies $\{n_k+\beta\}$ and a total of $N+K\beta$ observations.  To generalize to non-integer $\beta$, we observe that the likelihood function is $\cL(\pvec{p}) = \Pr(\{n_k\}|\pvec{p}) = \prod_k{p_k^{n_k}}$, and adding $\beta$ dummy observations of each letter yields a \emph{hedged} likelihood function
\begin{equation}\label{eq:ClassicalHedgedLikelihood}
\cL'(\pvec{p}) = \prod_k{p_k^{n_k+\beta}} = \left(\prod_k{n_k^\beta}\right)\cL(\pvec{p}),
\end{equation}
whose maximum value is achieved by Eq. \ref{eq:AddBetaRule}.  When $\beta$ is not an integer, the hedged likelihood (Eq. \ref{eq:ClassicalHedgedLikelihood}) remains well-defined, and the ``add $\beta$'' rule still maximizes it.

%\section{Quantum state estimation}

\textbf{Quantum Hedging:}  The quantum analogue of a distribution $\pvec{p}$ is a $d\times d$ density matrix $\rho$.  It cannot be observed directly; observing a sample of $\rho$ requires choosing a particular measurement $\cM$.  Experimentalists often divide the samples into groups and measure $\cM_j$ on the $N_j$ samples in group $j$, but $\cL(\rho)$ depends only on observed events, \emph{not} the unobserved alternatives, so we may pretend that all $N$ samples were measured by $\cM = \bigcup_j{w_j\cM_j}$, where $w_j =\frac{\scriptscriptstyle N_j}{\scriptscriptstyle N}$.  $\cM$ corresponds to a \emph{POVM}, a set of positive operators $\{\op{E}_i\}$ summing to $\Id$, which determine the probability of outcome ``$i$'' as
\begin{equation} \label{eq:BornRule}
\Pr(i) = \Tr[\rho\op{E}_i]. 
\end{equation}
The frequencies $\{n_i\}$ thus provide information about $\rho$.  Interpreting this information is the central problem of quantum state estimation.

The oldest and simplest procedure, linear inversion tomography \cite{VogelPRA89}, is based on Eq. \ref{eq:NaiveRule}.  Inverting Born's Rule (Eq. \ref{eq:BornRule}) yields an estimate $\rhotomo$ satisfying
\begin{equation} \label{eq:Tomography}
\Tr[\rhotomo\op{E}_i] = \frac{n_i}{N}\mathrm{\ for\ }i=1\ldots m.
\end{equation}
If these equations are overcomplete, $\rhotomo$ is chosen by least-squares fitting.  Frequently, some of $\rhotomo$'s eigenvalues are negative -- a serious problem, for they represent probabilities.  This occurs because linear inversion is blind to the shape of the space of quantum states (which assign probabilities to \emph{all} measurements).  It tries to fit data from a \emph{single} POVM $\cM$, and happily assigns negative probabilities for measurements that weren't performed.

The usual fix for this problem is MLE \cite{HradilPRA97}.  A likelihood function is derived from the data,
\begin{equation}\label{eq:QuantumLikelihood}
\cL(\rho) = \Pr(\{n_i\}|\rho) = \prod_i{\Tr[\rho \op{E}_i]^{n_i}},
\end{equation}
and we assign the $\rhohat$ that maximizes it.  Maximization over \emph{all} trace-1 Hermitian matrices yields $\rhotomo$ (just as in the classical case), but restricting to $\rho\geq0$ yields a non-negative $\rhoMLE$.

However, $\rhoMLE$ can still assign zero probabilities -- just like its classical counterpart (Eq. \ref{eq:NaiveRule}).  If $\rhotomo$ is not strictly positive, $\rhoMLE$ will have at least one zero eigenvalue \cite{RBK06b}, so this is rather common.  Moreover, the zero probabilities in $\rhoMLE$ are less justified than those in $\pMLE$, because they generally correspond to a measurement outcome $\proj{\psi}$ that is not an element of the measured POVM, and could never have appeared.  In contrast, Eq. \ref{eq:NaiveRule} assigns $p_k=0$ only when ``$k$'' has been given $N$ chances to appear and (so far) has not.  So although $\rhoMLE$ may be the right estimator for \emph{some} task, its zero eigenvalues represent a level of confidence that is implausible and (for predictive tasks like gambling and compression) catastrophic.  \emph{Prediction demands a hedged estimator.}

Bayesian mean estimators are hedged, and with suitable priors they have extremely good predictive behavior \cite{RBK06b}.  However, for quantum estimation there are no closed-form solutions, and numerical integration is hard.  This is unfortunate, because Bayes estimators for classical probabilities work very well.  They yield ``add $\beta$'' rules when applied to Dirichlet-$\beta$ priors, and Dirichlet priors are well motivated.  Jeffreys' prior ($\beta=\frac12$) yields asymptotically minimax-optimal estimators for data compression \cite{ClarkeJSPI94}, Krichevskiy showed that ``add $0.50922\ldots$'' outperforms all other rules for predicting the next letter \cite{KrichevskiyIEEE98}, and Braess et al \cite{BraessLNCS02} pointed out that $\beta\approx1$ generally works well because large-$N$ behavior depends only weakly on $\beta$.  

\begin{figure*}[th]
\begin{tabular}{lll}
\hspace{-0.2in}\includegraphics[width=\TPW]{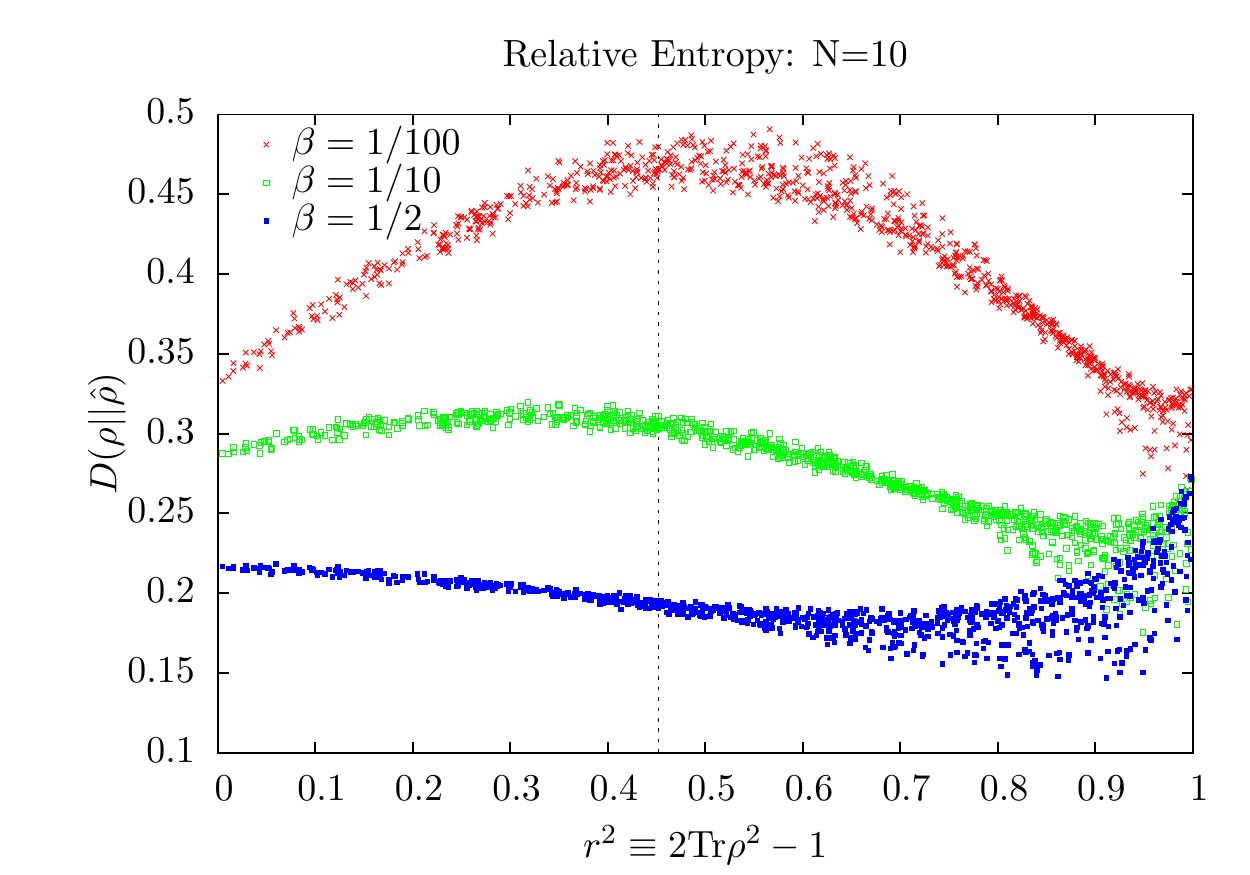} & \hspace{-0.2in}\includegraphics[width=\TPW]{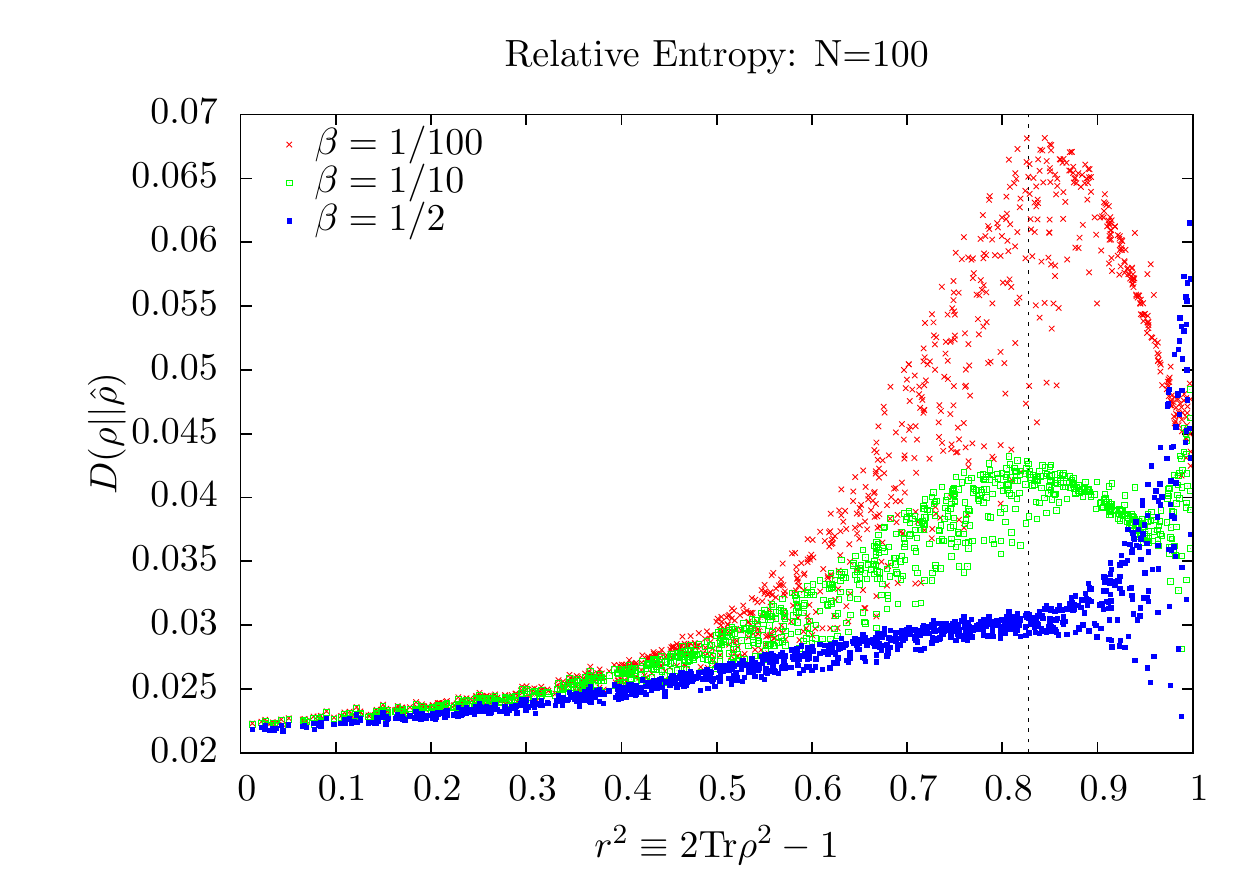} & \hspace{-0.2in}\includegraphics[width=\TPW]{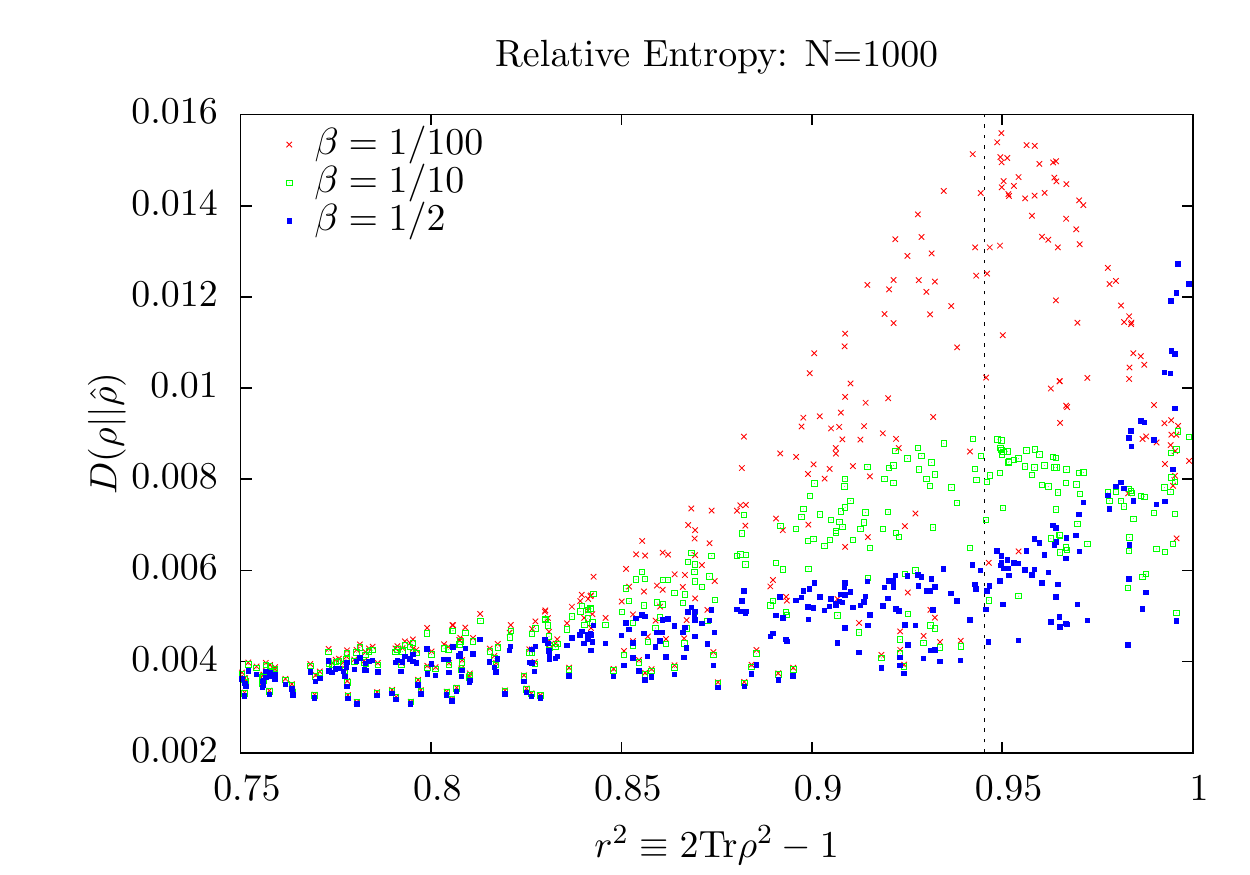}
\end{tabular}
\caption{\textbf{Methodology}: $10^3$ single-qubit states $\rho_\mathrm{true}$ were selected at random from the Hilbert-Schmidt (``flat'') measure on the Bloch sphere.  For each state, $10^3$ separate datasets were generated, each consisting of $3N$ ($N=10,100,1000$) measurements divided among the three Pauli operators.  HMLE estimates (with several $\beta$ values) were calculated.  For each $\rho_{\mathrm{true}}$, relative entropy error was averaged over all $10^3$ datasets. \textbf{Results}:  Error is strongly correlated with $r^2 = \frac12(1+\Tr\rhotrue^2)$.  There are three regimes, separated by $1-r^2\approx\sqrt{3/N}$ (dotted line).  \textbf{(1)} for mixed states with $1-r^2\gg\sqrt{3/N}$, accuracy increases slightly with the amount of hedging (quantified by $\beta$). \textbf{(2)} for slightly mixed states with $1-r^2 \approx \sqrt{3/N}$, accuracy improves substantially with hedging, but only up to $\beta\approx\frac12$. \textbf{(3)} for nearly-pure states with $1-r^2 \ll \sqrt{3/N}$, a small amount of hedging improves accuracy, but higher $\beta$ increases error, and the optimal $\beta$ decreases with $N$.}
\label{fig:QubitKL}
\end{figure*}

This suggests adapting ``add $\beta$'' to quantum state estimation (independent of Bayesian arguments).  However, obvious methods like adding dummy counts don't work.  Suppose we estimate a qubit source by measuring $\sigma_x$, $\sigma_y$, and $\sigma_z$ ten times each, and -- by unlikely chance -- all the outcomes are $+1$.  $\rhotomo$
%$$\rhotomo = \left(\begin{array}{cc} 1 & \frac{1+i}{2} \\ & \\ \frac{1-i}{2} & 0 \end{array}\right),$$
lies well outside the Bloch sphere, and $\rhoMLE$ is the projector onto its largest eigenvector.  Now, if we add $\beta=1$ dummy counts, $\rhotomo$
%$$\rhotomo' = \left(\begin{array}{cc} \frac{11}{12} & \frac{5(1+i)}{12} \\ & \\ \frac{5(1-i)}{12} & \frac{1}{12} \end{array}\right),$$
is still outside the Bloch sphere, and $\rhoMLE'$ is unchanged!

The underlying problem is that MLE tries to fit the observed data, with no consideration of unobserved measurements -- but the resulting quantum state makes predictions about those unobserved measurements.  Adding dummy data works in the classical case because there are only $K$ different events that can be observed \emph{or} predicted, so by adding a dummy observation of each one, we rule out the possibility of assigning $p_k=0$ to any event.  A quantum state assigns probabilities to infinitely many different events (measurement outcomes), and a finite set of dummy observations cannot bound all of these probabilities away from zero.

Instead, HMLE modifies the likelihood function directly, multiplying it by a unitarily invariant hedging function (Eq. \ref{eq:HedgingFunction}) that is independent of what POVM was measured.  This modification is directly analogous to the one generated by dummy counts in Eq. \ref{eq:ClassicalHedgedLikelihood}, because $\det(\rho)$ is the product of $\rho$'s eigenvalues.  In both cases, hedging makes very small probabilities less attractive, steering the maximum of $\cL'(\cdot)$ away from boundaries.  When the data are all drawn from a single classical basis (i.e., $\cM$ is a projective measurement), HMLE reproduces the ``add $\beta$'' rule exactly: if outcome $\proj{k}$ was observed $n_k$ times, then the HMLE estimate is
\begin{equation}
\rhoHMLE = \sum_k{\frac{n_k+\beta}{N+K\beta}\proj{k}}.
\end{equation}
Eq. \ref{eq:HedgingFunction} is the only measurement-independent smooth modification of $\cL(\rho)$ that yields ``add $\beta$'' for every basis (see Appendix B of the \texttt{arxiv.org} version for proof).

%\section{Performance}

\textbf{Performance:}  The point of HMLE is to give more accurate estimates than MLE.  ``Accuracy'' depends on the measure of error, but HMLE is motivated by the idea that a state should be \emph{predictive}, and predictive tasks (e.g., data compression and gambling) suggest that quantum relative entropy is a good measure of inaccuracy.
%\begin{equation}
%D(\rhotrue|\rhohat) \equiv \Tr\rhotrue\log\rhotrue - \Tr\rhotrue\log\rhohat.
%\end{equation}
This is a bit unfair to MLE.  If $\rhohat$ is rank-deficient on $\rhotrue$'s support, then $D(\rhotrue||\rhohat) = \infty$.  Since every true $\rhotrue$ has \emph{some} nonzero probability of serving up measurement results that yield a rank-deficient $\rhoMLE$, the expected value of $D(\rhotrue|\rhoMLE)$ is always infinite.  What we \emph{can} do is compare different hedging parameters.   Figure \ref{fig:QubitKL} shows relative-entropy error for $\beta = 10^{-2},10^{-1},\frac12$, applied to a single qubit measured $N=10,10^2,10^3$ times in each of the Pauli bases.  

The error depends on $\rhotrue$, most strongly on its radial coordinate $r = \sqrt{\frac{1+\Tr\rhotrue^2}{2}}$.  Three regimes are evident.  \textbf{(1)} For highly mixed states ($1-r^2\gg\sqrt{3/N}$), where MLE rarely yields rank-deficient estimates, accuracy increases slowly with $\beta$.  \textbf{(2)}  For slightly mixed states ($1-r^2\approx\sqrt{3/N}$), where MLE frequently yields a zero eigenvalue, accuracy improves dramatically with increased $\beta$, up to $\beta\approx1/2$.  \textbf{(3)}
Nearly-pure states ($1-r^2\ll\frac{3}{\sqrt{N}}$) display unexpected and complex behavior.  Optimal accuracy is achieved by a very small amount of hedging that decreases with $N$ as $\beta_{\mathrm{optimal}} \approx \frac{1}{2\sqrt{N}}$.  Beyond this point, more hedging leads to greater inaccuracy -- most noticeably for pure states.

\begin{figure}
\begin{tabular}{ll}
\hspace{-0.2in}\includegraphics[width=\HCW]{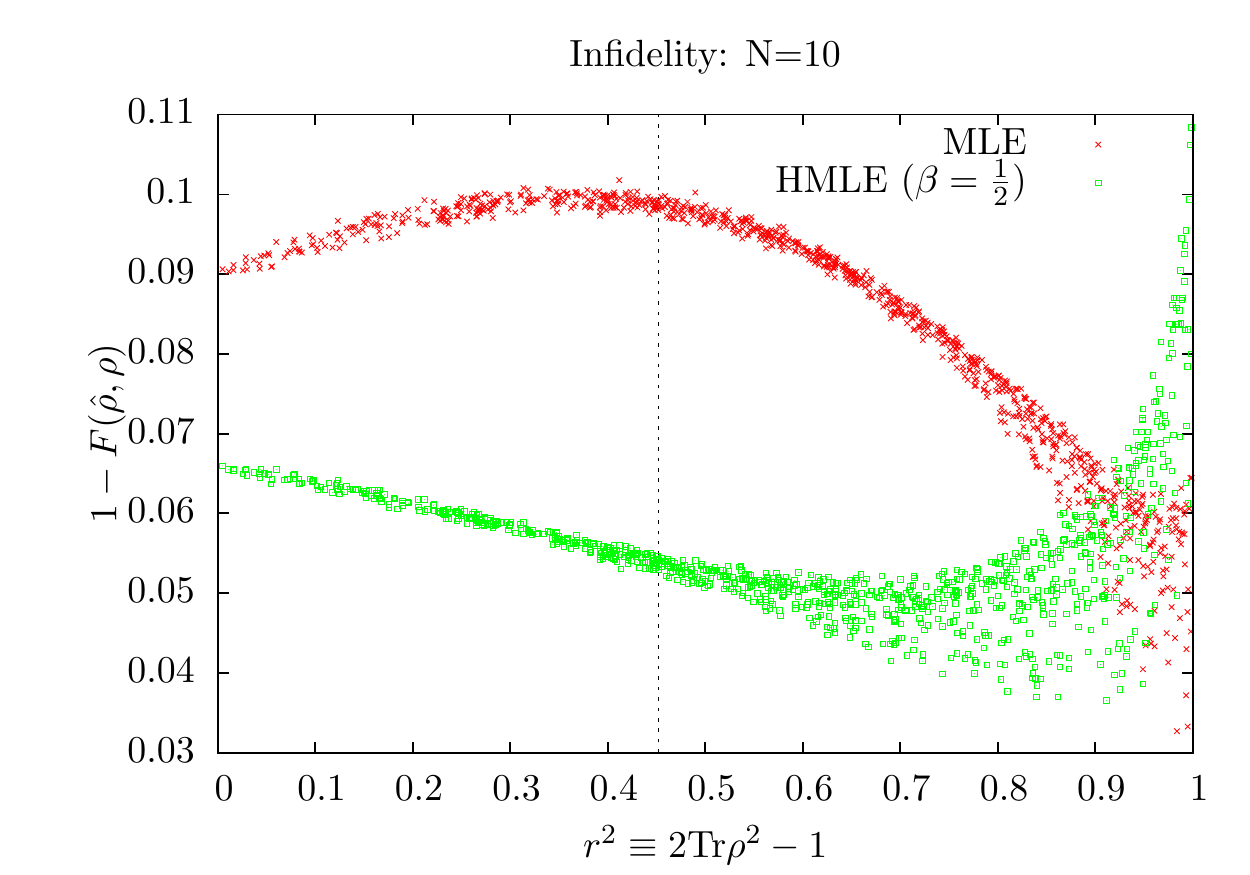} & \hspace{-0.2in} \includegraphics[width=\HCW]{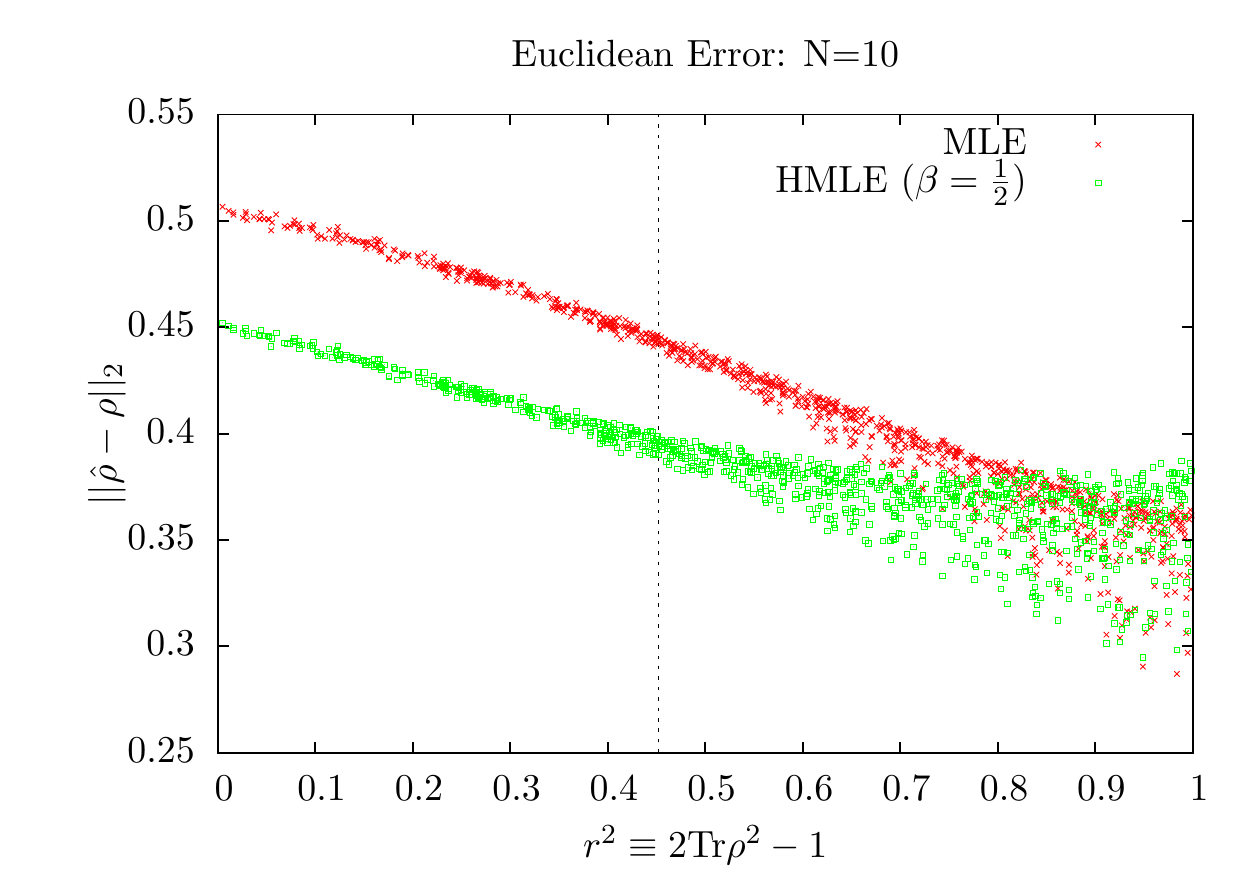} \\

\hspace{-0.2in}\includegraphics[width=\HCW]{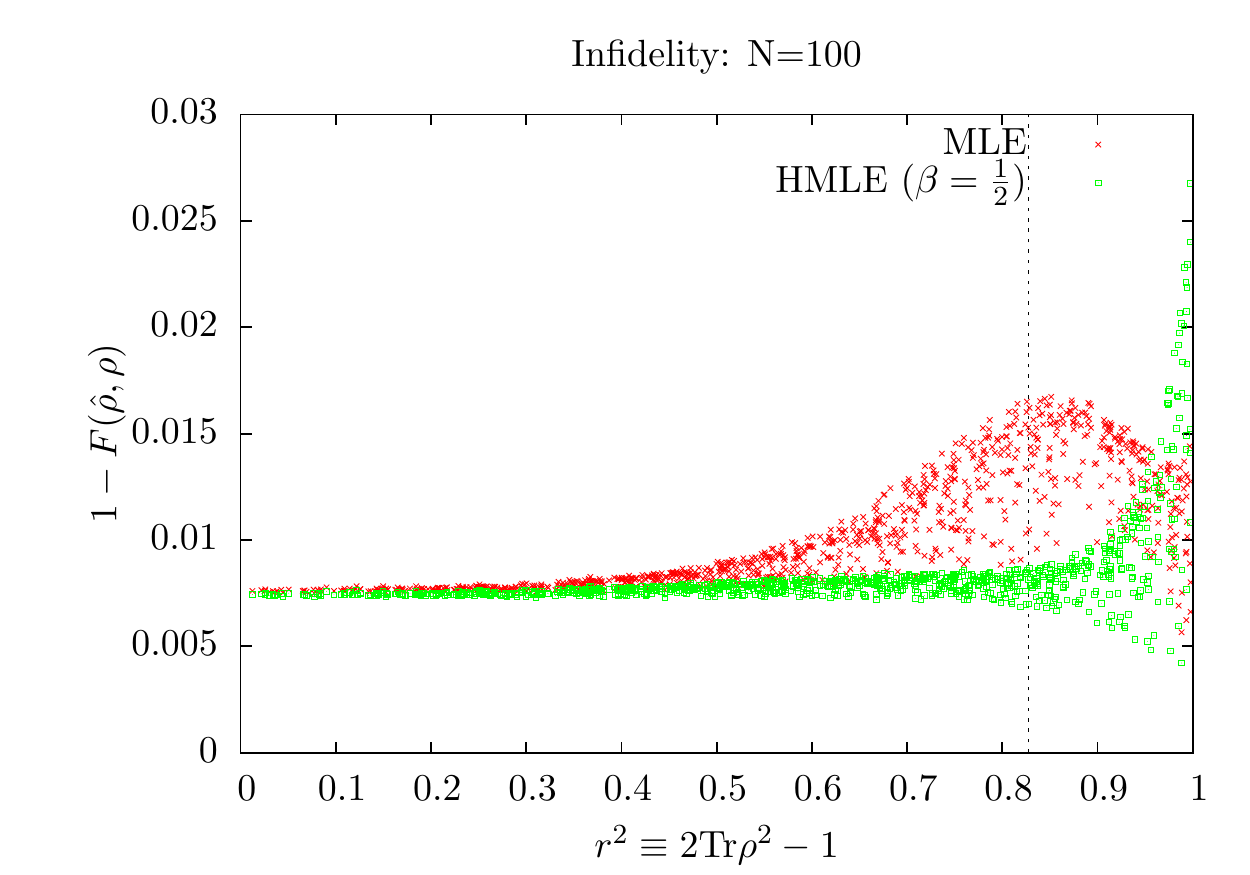} & \hspace{-0.2in}\includegraphics[width=\HCW]{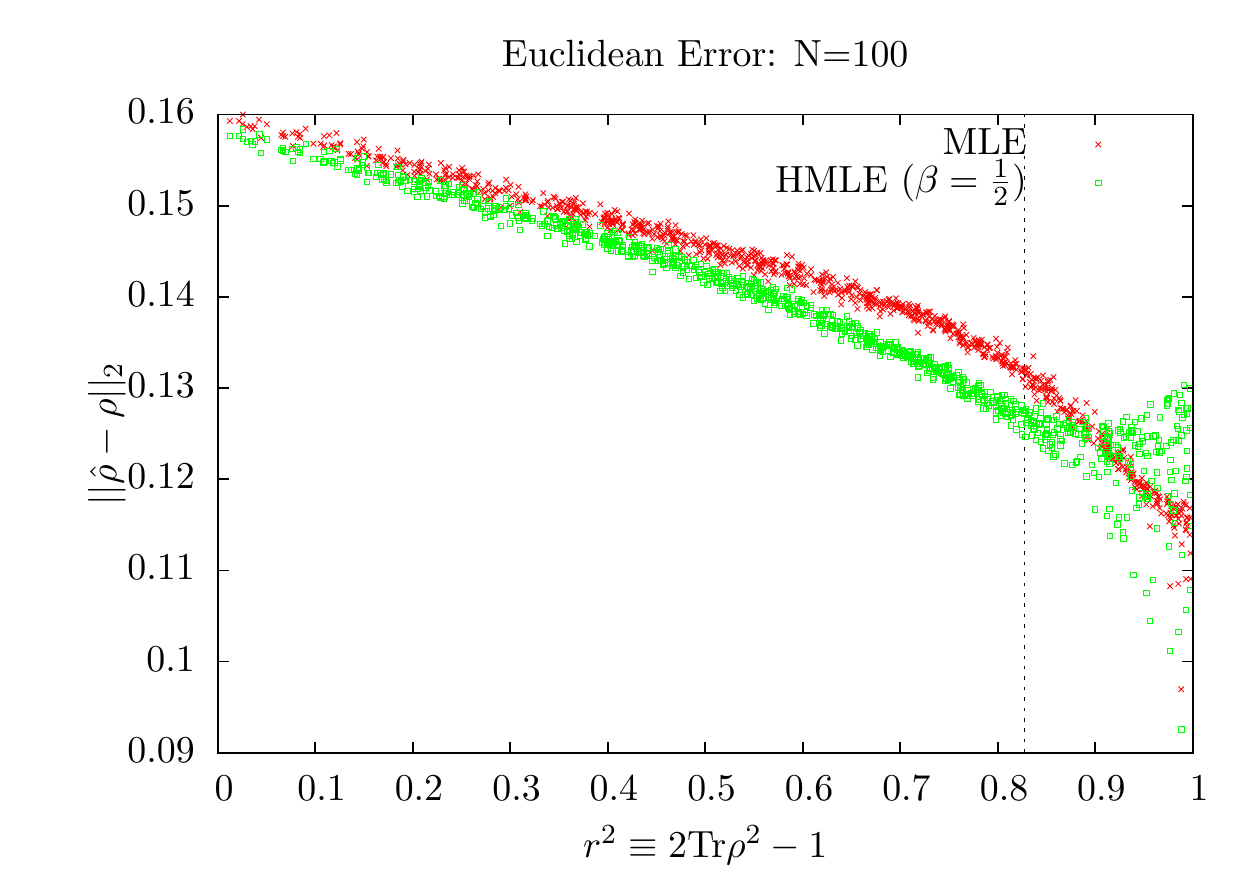} \\

\hspace{-0.2in}\includegraphics[width=\HCW]{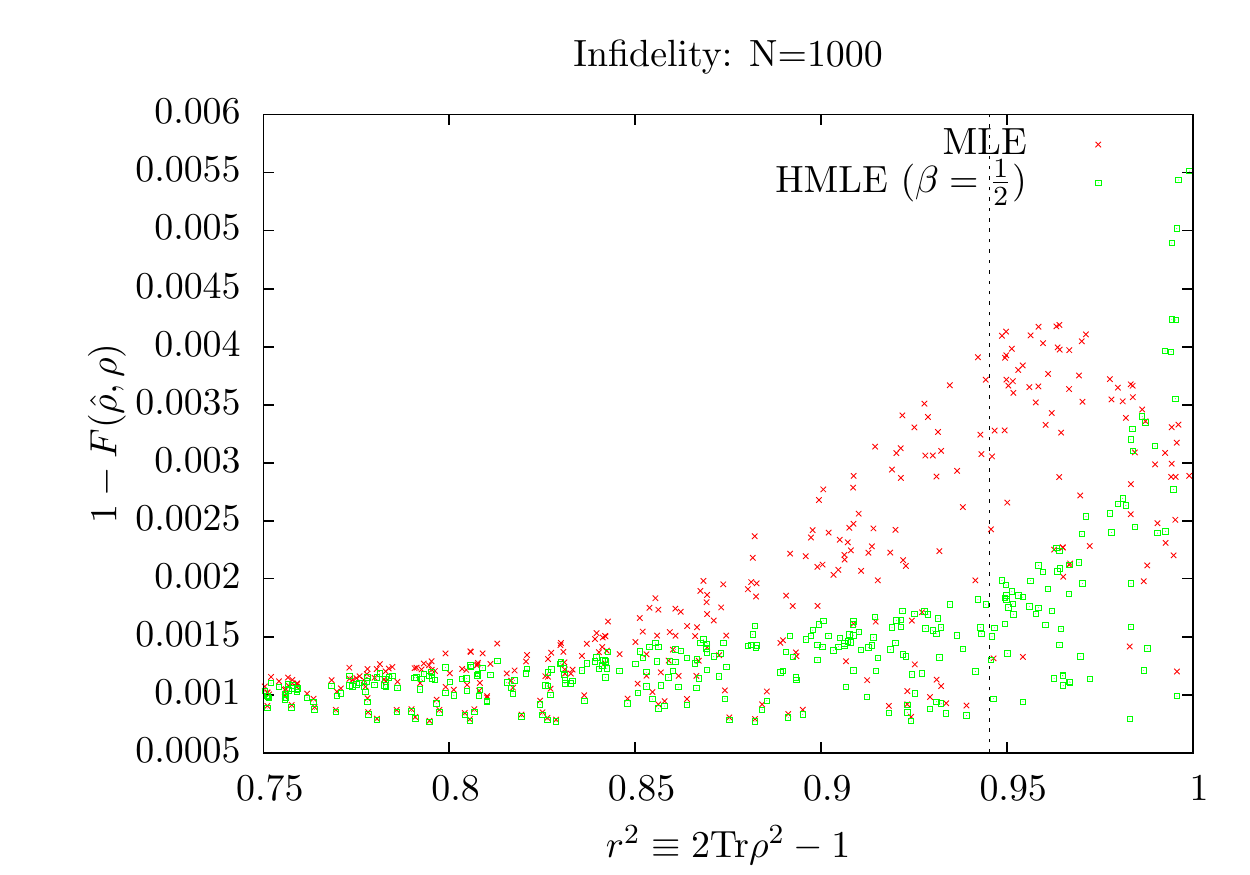} & \hspace{-0.2in}\includegraphics[width=\HCW]{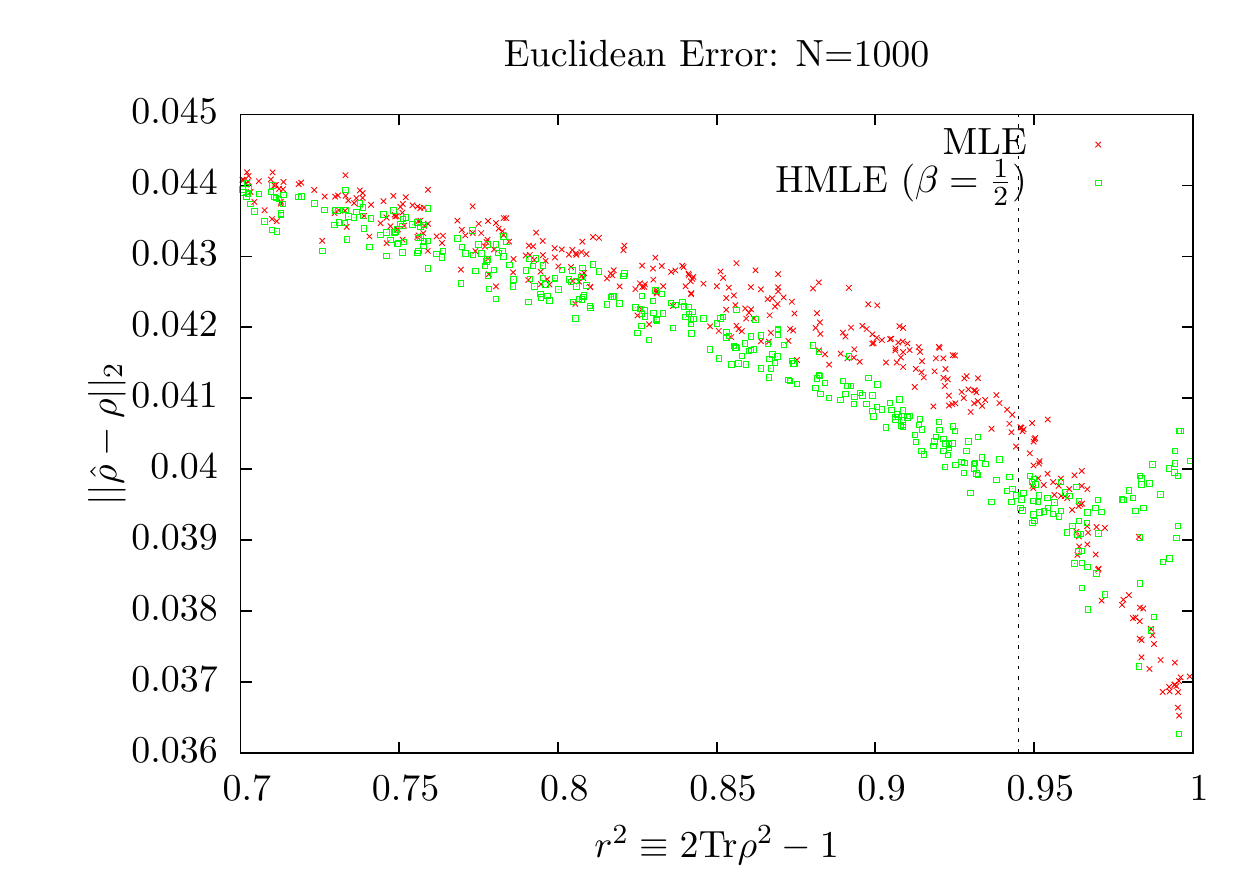} \\
\end{tabular}
\caption{\textbf{Methodology}: See Fig. \ref{fig:QubitKL}.  MLE and HMLE (with several $\beta$ values) estimates were calculated, and for each $\rho_{\mathrm{true}}$, Euclidean distance and infidelity were averaged over all datasets.  (Note that the trace and Euclidean distances are equivalent for qubits: $||\rhohat-\rho||_1 = \sqrt2||\rhohat-\rho||_2$.) \textbf{Results}:  The same three regimes are apparent as in Fig. \ref{fig:QubitKL}. For highly mixed states, hedging provides a small but consistent improvement.  Slightly mixed states see substantial improvement from hedging.  For nearly-pure states, hedging decreases accuracy regardless of $\beta$.  However, nearly-pure states are typically estimated with greater accuracy than more mixed states, so the best overall performance is achieved by hedging.  $\beta=0.25\ldots1$ seems to be optimal; for $\beta>1$, the error in pure state estimation outweighs benefits for mixed states.}
\label{fig:QubitErrors}
\end{figure}

Other error metrics include Euclidean distance ($\sqrt{\Tr[(\rhotrue-\rhohat)^2]}$), infidelity ($1-\left(\Tr\sqrt{\sqrt{\rho}\sigma\sqrt{\rho}}\right)^2$), and trace distance ($\Tr|\rho-\sigma|$) \cite{NielsenBook00}.  Each of these metrics has its purpose, but none of them are particularly appropriate for comparing $\hat\rho$ to $\rhotrue$.  Nonetheless, since they are widely used, Fig. \ref{fig:QubitErrors} illustrates their behavior for MLE and HMLE, applied to a single qubit measured in the Pauli bases.  Both Euclidean/trace-norm distance (they are equivalent for qubits) and infidelity show the same basic behavior.  For nearly pure states, MLE is more accurate.  For highly mixed states, HMLE improves accuracy slightly.  The biggest improvement comes in the intermediate regime where $O(1/N) < 1-r^2 < O(1/\sqrt{N})$.  These states are not quite pure, but close enough that MLE yields rank-deficient estimates a substantial fraction of the time.  In this regime, hedging provides substantial improvement.  So, even though HMLE is not designed to maximize fidelity or trace-distance, it improves on MLE for all but the purest states.

%\section{Discussion and Conclusions}

\textbf{Discussion:}  There is nothing sacred about the maximum of $\cL(\rho)$, but $\rhoHMLE$ should not be \emph{significantly} less likely than $\rhoMLE$.  Likelihood measures plausibility, and if $\cL(\rhoHMLE)$ is almost as large as $\cL(\rhoMLE)$, then $\rhoHMLE$ is almost as plausible as $\rhoMLE$.  If they have identical properties, we may as well pick the more plausible one -- but if $\rhoHMLE$ is substantially different in some way, then it should be considered on its merits unless $\rhoMLE$ has significantly higher likelihood.  We have already seen that the HMLE estimate has substantially different properties, so let's confirm that it is not significantly less likely.

Consider the classical case.  If $n_k=0$, then $\pMLE$ assigns $\hat{p}_k=0$.  But it's equally plausible that $p_k>0$, since if $p_k < \frac{1}{N}$ then ``$k$'' probably won't appear in the first $N$ samples.  The likelihood function bears this out:  the \emph{most} likely state sets $\hat{p}_k = 0$, but nearby states with nonzero $p_k$ have almost the same likelihood.  If $\pHMLE$ assigns $p'_k = \frac{\beta}{N}$ and $p'_j = \left(1-\frac{\beta}{N}\right)\frac{n_j}{N}$ for $j\neq k$, then
\begin{equation} \label{eq:LikelihoodRatio}
\frac{\cL(\pHMLE)}{\cL(\pMLE)} = \left(1-\frac{\beta}{N}\right)^N \approx e^{-\beta}.
\end{equation}
Likelihood ratios between $e^{-1}$ and $e$ are ``barely worth mentioning'' \cite{JeffreysBook98}, so if $\beta<1$, then $\pHMLE$ is essentially as plausible as $\pMLE$.  Actually, $\pMLE$ comprises $K-1$ independent parameters, and in this case likelihood ratios between $e^{-K}$ and $e^{K}$ are insignificant.  [Typically, $\cL(\pvec{p}_{\mathrm{true}})\approx e^{-K}\cL(\pMLE)$, so tighter significance criteria would reject the true state.]  If $\pMLE$ assigns zero probability to $M<K$ different events, and $\pHMLE$ hedges all $M$ of them, then the argument leading to Eq. \ref{eq:LikelihoodRatio} gives a likelihood ratio of $e^{-M\beta}$, which is not significant.

For quantum HMLE, it's possible to show the same result for the HMLE estimate $\rhoHMLE$:
\begin{equation}
\frac{\cL(\rhoHMLE)}{\cL(\rhoMLE)} \geq e^{-d\beta}.
\end{equation}
The proof is a bit long, and can be found in Appendix A of the \texttt{arxiv.org} version.

\textbf{Conclusions:}  Hedging is a simple, well-motivated solution to the zero eigenvalue problem.  It is also easy to implement -- unlike, for instance, Bayesian techniques (which tend to be an order of magnitude harder to calculate).  HMLE can be implemented by a near-trivial change to any MLE routine. Because the hedged likelihood goes smoothly to zero near the boundary, no explicit positivity constraint is needed.  So in fact it may be \emph{easier} than MLE, as simple gradient-crawling methods should work (though care is necessary for small $\beta$, where the boundary roll-off becomes sharper).

Because $\rhoHMLE$ is always full-rank, it can safely be used for predictive tasks like gambling and data compression.  HMLE works well for qubits, providing improved accuracy by almost all metrics.  Further studies will reveal how well it works for larger systems.  Pure states are best estimated using very small values of $\beta$, and in general the optimal value of $\beta$ is not clear.  This contrast with the classical case, where $\beta\approx\frac12$ is known to be asymptotically optimal, suggests that alternative hedging functions (which do not correspond to ``add $\beta$'') may work better for quantum estimation.

\textbf{Acknowledgments:}  This paper owes much to discussions with Alexei Gilchrist, Daniel James, Jordan LaPointe, and Rob Spekkens.  I am supported by the Government of Canada through Industry Canada and by the Province of Ontario through the Ministry of Research \& Innovation.

\appendix

\section{Appendix A}

The point of this section is to demonstrate that the \emph{hedged} maximum likelihood estimate $\rhoHMLE$ is never significantly less plausible than the MLE estimate, i.e.
\begin{equation*}
\frac{\cL(\rhoHMLE)}{\cL(\rhoMLE)} \geq e^{-d\beta}.
\end{equation*}

The MLE estimate maximizes the log-likelihood ($l(\rho) \equiv \log\cL(\rho)$), while the HMLE estimate maximizes the hedged log-likelihood ($l'(\rho) = l(\rho) + \log h(\rho)$, where $h(\rho)$ is given in Eq. \ref{eq:HedgingFunction}). Both are convex functions on a convex subset of $\reals^{d^2-1}$.  It is convenient to think of $-l(\rho)$ and $-l'(\rho)$ as potential energy functions, and of $\rhoMLE$ and $\rhoHMLE$ as the corresponding equilibrium states.  In this picture, the gradients $\grad l(\rho)$ and $\grad \log h(\rho)$ are force fields (which balance perfectly at $\rhoHMLE$), and the logarithm of the likelihood ratio
$$\log\left(\frac{\cL(\rhoMLE)}{\cL(\rhoHMLE)}\right) = l(\rhoMLE) - l(\rhoHMLE) = \Delta l$$
is the amount of work done by $\grad \log h(\rho)$ by adiabatically changing the equilibrium from $\rhoMLE\to\rhoHMLE$.

Because $h(\rho)$ depends only on $\rho$'s eigenvalues, the corresponding ``force''
$$\grad \log h(\rho) = \beta\rho^{-1}$$
is orthogonal to unitary rotations, and acts only on the spectrum of $\rho$.  Furthermore, while it diverges at the boundary, it becomes rapidly and monotonically weaker away from the boundary.  So, although it inexorably forces $\rhohat$ off the boundary, it does not necessarily push it very far.

Let us imagine that the hedging parameter (denoted $\beta'$) is adiabatically increased from zero to $\beta$.  For each $\beta'$, there is an equilibrium $\rhohat_{\beta'}$.  Increasing $\beta'$ by $\diff \beta$ shifts it a distance $\diff \rhohat$ and does work
$$\diff l = -\op\grad l\cdot\diff\rhohat = \op\grad \log h|_{\beta'} \cdot \diff \rhohat.$$
Integrating $\op\grad \log h|_{\beta'} \cdot \diff \rhohat$ along the entire path yields $\Delta l$.  Since $\grad\log h$ is orthogonal to unitary changes in $\rhohat$, the integral is only sensitive to motion within the eigenvalue simplex, so
$$\op\grad \log h|_{\beta'} \cdot \diff \rhohat = \sum_k{\pdiff{\log h}{\lambda_k}\diff \lambda_k}.$$
It's tempting to evaluate this directly as
$$\Delta l = \int_{\rho=\rhoMLE}^{\rhoHMLE}{\beta'\sum_k{\frac{\diff\lambda_k}{\lambda_k}}},$$
but $\beta'$ changes with $\rho$.  Instead, we upper-bound the integral by observing that $-l(\rho)$ and $-\log h(\rho)$ are both concave, so their second derivatives are strictly positive.  As $\rhohat_{\beta'}$ moves \emph{away} from the maximum of $l(\rho)$ and \emph{toward} that of $\log h(\rho)$, the components of $-\op\grad l$ and $\op\grad\log h$ parallel to $\diff\rhohat$ are strictly increasing.  Thus, substituting $\op\grad\log h$ evaluated at $\rhoHMLE$ into the integral yields an upper bound.  Defining the eigenvalues of $\rhoMLE$ as $\lambda^0_k$ and those of $\rhoHMLE$ as $\lambda^f_k$,
\begin{eqnarray*}
\Delta l &\leq& \int_{\rho=\rhoMLE}^{\rhoHMLE}{\beta\sum_k{\frac{\diff\lambda_k}{\lambda_k}}} \\
&=& \beta\sum_{k=1}^d{\frac{\lambda^f_k-\lambda^0_k}{\lambda^f_k}} \\
&\leq& \beta d.
\end{eqnarray*}
This means that $\cL(\rhoHMLE)$ is at least $e^{-d\beta}\cL(\rhoMLE)$, so for $\beta<1$ it is not significantly less plausible.

This does not necessarily mean that $\rhoHMLE$ is close to $\rhoMLE$.  When $\cL(\rho)$ is nearly flat, hedging can cause substantial deflection -- precisely because there is no gradient in $\cL$ to oppose it.  When $\cL$ is sharply peaked around $\rhoMLE$, hedging has comparatively little effect.

\section{Appendix B}

The point of this section is to show that the hedging function given in Eq. \ref{eq:HedgingFunction},
$$h(\rho) = \det(\rho)^\beta,$$
is the only smooth hedging function that reproduces the ``add $\beta$'' rule for measurements in \emph{any} single orthonormal basis.  That is, when $N$ i.i.d. $d$-dimensional quantum systems all have been measured in a single basis denoted $\{\ket{0}\ldots\ket{d-1}\}$, and outcome $\proj{k}$ has appeared $n_k$ times, the maximum of the hedged likelihood $\cL' = h(\rho)\cL(\rho)$ should be
\begin{equation} \label{eq:BetaHMLE}
\rhoHMLE = \sum_k{ \frac{n_k + \beta}{N+d\beta} \proj{k} }.
\end{equation}

First, consider hedging according to Eq. \ref{eq:HedgingFunction}.  The hedged likelihood is
$$\cL'(\rho) = \det(\rho)^\beta\prod_k{\braopket{k}{\rho}{k}^{n_k}}.$$
It's equally valid (and more convenient) to maximize its logarithm,
$$\log\cL'(\rho) = \beta\log\det(\rho) + \sum_k{n_k\log\braopket{k}{\rho}{k}}.$$
This function's gradient thus has two components, one from the likelihood and one from the hedging function.  The likelihood depends \emph{only} on the diagonal elements of $\rho$, so its gradient is orthogonal to off-diagonal variations.  The hedging function is unitarily invariant, so its gradient is orthogonal to unitary rotations.  If we vary only over diagonal matrices $\rho = \sum_k{p_k\proj{k}}$, then this problem reduces to the classical one and it's easy to show that Eq. \ref{eq:BetaHMLE} is the maximum.  Furthermore, this is a local maximum (with respect to \emph{all} variations), because the gradient of $h(\rho)$ is orthogonal to unitary rotations and therefore locally orthogonal to off-diagonal variations.  This is also a global maximum, because $\log\cL'(\rho)$ is a convex function.  Thus, Eq. \ref{eq:BetaHMLE} maximizes the hedged likelihood.

Now, consider some other hedging function $h'(\rho)$.  If $h'(\rho)$ is not unitarily invariant, then there exists some point $\rho$ such that, in the neighborhood of $\rho$, the gradient of $h'(\rho)$ is not orthogonal to unitary rotations.  Suppose that the measured basis $\{\ket{k}\}$ is the diagonal basis of $\rho$, and the measured frequencies $n_k$ are such that $\rhoHMLE$ (given by Eq. \ref{eq:BetaHMLE}) is in the neighborhood of $\rho$.  Then, at the point $\rhoHMLE$, the gradient of $\cL(\rho)$ is orthogonal to off-diagonal variations, but the gradient of $h'(\rho)$ is not.  This means that the gradient of the hedged likelihood does not vanish, and thus its maximum cannot coincide with Eq. \ref{eq:BetaHMLE}.

If $h'(\rho)$ \emph{is} unitarily invariant, then for every measured basis, the maximization can safely be restricted to diagonal matrices $\rho = \sum_k{p_k\proj{k}}$, and it reduces to a classical problem, maximizing
$$\log\cL'(\pvec{p}) = \log h'(\pvec{p}) + \log\cL(\pvec{p}).$$
To reproduce the ``add $\beta$'' rule, the gradient of $\log\cL'$ must vanish -- for all $\{n_k\}$ -- at $\pHMLE = \left\{\frac{n_k+\beta}{N}\right\}$.  This implies
$$\grad\log h'|_{\pHMLE} = - \grad\log\cL|_{\pHMLE},$$
and since this condition is automatically satisfied by Eq. \ref{eq:HedgingFunction},
$$\grad\log h'|_{\pHMLE} = \grad\log h|_{\pHMLE}$$
at every point $\pHMLE = \left\{\frac{n_k+\beta}{N}\right\}$ for any $\{n_k\}$.  These points are dense in the simplex on which $h'(\pvec{p})$ is defined.  This means that $h(\rho) = h'(\rho)$, since smooth functions whose derivatives agree at a dense set of points are identical.

\bibliographystyle{apsrev}
\bibliography{HMLE.bib,../../bib/quantum,../../bib/math,../../bib/estimation,../../bib/RBK}

\begin{thebibliography}{15}
\expandafter\ifx\csname natexlab\endcsname\relax\def\natexlab#1{#1}\fi
\expandafter\ifx\csname bibnamefont\endcsname\relax
  \def\bibnamefont#1{#1}\fi
\expandafter\ifx\csname bibfnamefont\endcsname\relax
  \def\bibfnamefont#1{#1}\fi
\expandafter\ifx\csname citenamefont\endcsname\relax
  \def\citenamefont#1{#1}\fi
\expandafter\ifx\csname url\endcsname\relax
  \def\url#1{\texttt{#1}}\fi
\expandafter\ifx\csname urlprefix\endcsname\relax\def\urlprefix{URL }\fi
\providecommand{\bibinfo}[2]{#2}
\providecommand{\eprint}[2][]{\url{#2}}

\bibitem[{\citenamefont{Nielsen and Chuang}(2000)}]{NielsenBook00}
\bibinfo{author}{\bibfnamefont{M.~A.} \bibnamefont{Nielsen}} \bibnamefont{and}
  \bibinfo{author}{\bibfnamefont{I.~L.} \bibnamefont{Chuang}},
  \emph{\bibinfo{title}{Quantum Computation and Quantum Information}}
  (\bibinfo{publisher}{Cambridge Press}, \bibinfo{year}{2000}).

\bibitem[{\citenamefont{Hradil}(1997)}]{HradilPRA97}
\bibinfo{author}{\bibfnamefont{Z.}~\bibnamefont{Hradil}},
  \bibinfo{journal}{Phys. Rev. A} \textbf{\bibinfo{volume}{55}},
  \bibinfo{pages}{R1561} (\bibinfo{year}{1997}).

\bibitem[{\citenamefont{James et~al.}(2001)\citenamefont{James, Kwiat, Munro,
  and White}}]{JamesPRA01}
\bibinfo{author}{\bibfnamefont{D.~F.~V.} \bibnamefont{James}},
  \bibinfo{author}{\bibfnamefont{P.~G.} \bibnamefont{Kwiat}},
  \bibinfo{author}{\bibfnamefont{W.~J.} \bibnamefont{Munro}}, \bibnamefont{and}
  \bibinfo{author}{\bibfnamefont{A.~G.} \bibnamefont{White}},
  \bibinfo{journal}{Physical Review A} \textbf{\bibinfo{volume}{64}},
  \bibinfo{pages}{052312 } (\bibinfo{year}{2001}).

\bibitem[{\citenamefont{Hradil et~al.}(2004)\citenamefont{Hradil, Reh\'{a}cek,
  Fiurasek, and Jezcaronek}}]{HradilLNP04}
\bibinfo{author}{\bibfnamefont{Z.}~\bibnamefont{Hradil}},
  \bibinfo{author}{\bibfnamefont{J.}~\bibnamefont{Reh\'{a}cek}},
  \bibinfo{author}{\bibfnamefont{J.}~\bibnamefont{Fiurasek}}, \bibnamefont{and}
  \bibinfo{author}{\bibfnamefont{M.}~\bibnamefont{Jezcaronek}}, in
  \emph{\bibinfo{booktitle}{Quantum state estimation}}, edited by
  \bibinfo{editor}{\bibfnamefont{M.~G.~A.} \bibnamefont{Paris}}
  \bibnamefont{and} \bibinfo{editor}{\bibfnamefont{J.}~\bibnamefont{Rehacek}}
  (\bibinfo{publisher}{Berlin, Germany : Springer-Verlag, 2004},
  \bibinfo{year}{2004}), vol. \bibinfo{volume}{649} of
  \emph{\bibinfo{series}{Lecture Notes In Physics}}, pp.
  \bibinfo{pages}{59--112}.

\bibitem[{\citenamefont{Haeffner et~al.}(2005)\citenamefont{Haeffner, Haensel,
  Roos, Benhelm, al~kar, Chwalla, Koerber, Rapol, Riebe, Schmidt
  et~al.}}]{HaeffnerNature05}
\bibinfo{author}{\bibfnamefont{H.}~\bibnamefont{Haeffner}},
  \bibinfo{author}{\bibfnamefont{W.}~\bibnamefont{Haensel}},
  \bibinfo{author}{\bibfnamefont{C.~F.} \bibnamefont{Roos}},
  \bibinfo{author}{\bibfnamefont{J.}~\bibnamefont{Benhelm}},
  \bibinfo{author}{\bibfnamefont{D.~C.} \bibnamefont{al~kar}},
  \bibinfo{author}{\bibfnamefont{M.}~\bibnamefont{Chwalla}},
  \bibinfo{author}{\bibfnamefont{T.}~\bibnamefont{Koerber}},
  \bibinfo{author}{\bibfnamefont{U.~D.} \bibnamefont{Rapol}},
  \bibinfo{author}{\bibfnamefont{M.}~\bibnamefont{Riebe}},
  \bibinfo{author}{\bibfnamefont{P.~O.} \bibnamefont{Schmidt}},
  \bibnamefont{et~al.}, \bibinfo{journal}{Nature}
  \textbf{\bibinfo{volume}{438}}, \bibinfo{pages}{643} (\bibinfo{year}{2005}).

\bibitem[{\citenamefont{Blume-Kohout}(2006)}]{RBK06b}
\bibinfo{author}{\bibfnamefont{R.}~\bibnamefont{Blume-Kohout}}
  (\bibinfo{year}{2006}), \bibinfo{note}{\texttt{quant-ph/0611080}}.

\bibitem[{\citenamefont{Lidstone}(1920)}]{LidstoneTFA20}
\bibinfo{author}{\bibfnamefont{G.~J.} \bibnamefont{Lidstone}},
  \bibinfo{journal}{Transactions of the Faculty of Actuaries}
  \textbf{\bibinfo{volume}{8}}, \bibinfo{pages}{182} (\bibinfo{year}{1920}).

\bibitem[{\citenamefont{Ristad}(1995)}]{Ristad95}
\bibinfo{author}{\bibfnamefont{E.}~\bibnamefont{Ristad}},
  \bibinfo{journal}{Arxiv preprint cmp-lg/9508012}  (\bibinfo{year}{1995}).

\bibitem[{\citenamefont{Cover and Thomas}(1991)}]{CoverBook91}
\bibinfo{author}{\bibfnamefont{T.~H.} \bibnamefont{Cover}} \bibnamefont{and}
  \bibinfo{author}{\bibfnamefont{J.~A.} \bibnamefont{Thomas}},
  \emph{\bibinfo{title}{Elements of Information Theory}}
  (\bibinfo{publisher}{Wiley-Interscience}, \bibinfo{year}{1991}).

\bibitem[{\citenamefont{Xie and Barron}(2000)}]{XieIEEE00}
\bibinfo{author}{\bibfnamefont{Q.}~\bibnamefont{Xie}} \bibnamefont{and}
  \bibinfo{author}{\bibfnamefont{A.}~\bibnamefont{Barron}},
  \bibinfo{journal}{IEEE Transactions on Information Theory}
  \textbf{\bibinfo{volume}{46}}, \bibinfo{pages}{431} (\bibinfo{year}{2000}).

\bibitem[{\citenamefont{Vogel and Risken}(1989)}]{VogelPRA89}
\bibinfo{author}{\bibfnamefont{K.}~\bibnamefont{Vogel}} \bibnamefont{and}
  \bibinfo{author}{\bibfnamefont{H.}~\bibnamefont{Risken}},
  \bibinfo{journal}{Phys. Rev. A} \textbf{\bibinfo{volume}{40}},
  \bibinfo{pages}{2847} (\bibinfo{year}{1989}).

\bibitem[{\citenamefont{Clarke and Barron}(1994)}]{ClarkeJSPI94}
\bibinfo{author}{\bibfnamefont{B.}~\bibnamefont{Clarke}} \bibnamefont{and}
  \bibinfo{author}{\bibfnamefont{A.}~\bibnamefont{Barron}},
  \bibinfo{journal}{Journal of Statistical Planning and Inference}
  (\bibinfo{year}{1994}).

\bibitem[{\citenamefont{Krichevskiy}(1998)}]{KrichevskiyIEEE98}
\bibinfo{author}{\bibfnamefont{R.}~\bibnamefont{Krichevskiy}},
  \bibinfo{journal}{IEEE Transactions on Information Theory}
  \textbf{\bibinfo{volume}{44}}, \bibinfo{pages}{296} (\bibinfo{year}{1998}).

\bibitem[{\citenamefont{Braess et~al.}(2002)\citenamefont{Braess, Forster,
  Sauer, and Simon}}]{BraessLNCS02}
\bibinfo{author}{\bibfnamefont{D.}~\bibnamefont{Braess}},
  \bibinfo{author}{\bibfnamefont{J.}~\bibnamefont{Forster}},
  \bibinfo{author}{\bibfnamefont{T.}~\bibnamefont{Sauer}}, \bibnamefont{and}
  \bibinfo{author}{\bibfnamefont{H.}~\bibnamefont{Simon}},
  \bibinfo{journal}{Lecture Notes in Computer Science} pp.
  \bibinfo{pages}{380--394} (\bibinfo{year}{2002}).

\bibitem[{\citenamefont{Jeffreys}(1998)}]{JeffreysBook98}
\bibinfo{author}{\bibfnamefont{H.}~\bibnamefont{Jeffreys}},
  \emph{\bibinfo{title}{Theory of Probability}} (\bibinfo{publisher}{Oxford
  University Press}, \bibinfo{year}{1998}).

\end{thebibliography}

\end{document}